\documentclass{amsart}
\usepackage{amsmath}
\usepackage{amssymb}

\newtheorem{thm}{Theorem}

\newcommand{\D}{\mathcal{D}}  

 \newcommand{\qtext}[1]{\quad\text{#1}\quad}
 \newcommand{\fa}{\qtext{for all}}
 \newcommand{\bb}{\begin{equation*}}
 \newcommand{\ee}{\end{equation*}}
 \newcommand{\bp}{\begin{proof}}
 \newcommand{\ep}{\end{proof}}

\begin{document}

\title[N-Body Problem of Electrodynamics]{The Existence and Uniqueness of  Solutions \\
 to N-Body Problem of Electrodynamics
}

\author{Victor M. Bogdan
}
\bigskip

\address{Department of Mathematics, McMahon Hall 207, CUA, Washington, D.C. 20064}

\email{bogdan@cua.edu, web fttp://faculty.cua.edu/bogdan}

%
%


\begin{abstract}
Given $n$ charges interacting with each other according
to Feynman's law. Let $(r_j(t),v_j(t))$ denote the
position and velocity of the charge $q_j.$ The list $y(t)$ of all such vectors is
called a trajectory.
A Lipschitzian trajectory $x(t), (t\le0),$ with continuous derivative, on which the velocities
do not exceed some limiting velocity $v<c,$ where $c$ denotes the speed of light,
is called an  initial trajectory.
A locally Lipschitzian trajectory $y(t)$
is called relativistically admissible
if the velocities on it stay below the speed of light $c.$
 The author constructs operators $\Phi_{j}$ of a trajectory whose values
$\Phi_j(y)(t)$ are linear transformations of $R^3$ into $R^3.$
A point $t=t_1$ on a trajectory $y$ is called singular
if either some of the charges collide at the time $t_1$
or the determinant is zero for at least one of the transformations
$\Phi_j(y)(t_1).$
 The main result is the following:
If $x(t)\,(t\le0)$ is an initial trajectory with  nonsingular point $t=0,$
then there exists a unique relativistically admissible
trajectory $y(t),$ defined for
$t$ in an interval $I\subset \langle 0,\infty),$
extending the initial trajectory $x(t)$
and having the following properties.
(1) No point $t$ on the trajectory $y$ is singular.
(2) The trajectory represents a unique solution of the
Newton-Einstein momentum-force system of equations
under Lorentz forces induced by electromagnetic field
in accord to Feynman's law for moving point
charges.
(3) The trajectory $y$ represents the maximal global solution
of the system.

\end{abstract}
%
\maketitle
\bigskip

Assume that we have a system of $n$ charges interacting with each other according
to Feynman's law for moving point charges%
Assume that $y_j(t)=(r_j(t),v_j(t))$ denotes the
position and velocity of the charge $q_j.$ The list $y(t)$ of all such vectors is
called a {\em trajectory} of the system.
Let $\D$ denote the differential operator $\frac{d }{d t}.$
Consider the Newton-Einstein system of equations with Lorentz forces
\begin{equation}
\label{n-body problem}
    \begin{split}
  \D p_j =&\ q_j\sum_{k\neq j}(E_{jk}+\frac{1}{c}\,v_j\times[ e_{jk}\times E_{jk}])
    \ (j=1,\ldots,n) \\
  E_{jk} =&\ \frac{q_k}{4\pi\epsilon_0}\Big[\frac{e_{jk}}{|r_{jk}|^2}+
  \frac{|r_{jk}|}{c}\D\frac{e_{jk}}{|r_{jk}|^2}+\frac{1}{c^2}\D^2e_{jk}\Big]
  \,(j\neq k)
  \end{split}
\end{equation}
where $E_{jk}$ represents the
intensity of the electric field acting onto the charge $q_j$ and generated by the charge
$q_k,$ in accord with Feynman's law for moving point charges.

The quantity $p_j$ denotes the relativistic momentum
of the charge $q_j.$
In the above formula $r_{jk}$ represents
the vector  starting at a point on the trajectory
of the charge $q_k,$ when the wave was emitted at time $t_{jk},$ and ending at the point
$r_j(t)$  at the time $t$ on the trajectory of the charge $q_j,$ when the wave arrived.
The vector $e_{jk}$ denotes
the unit vector in the direction of the vector $r_{jk}.$


A Lipschitzian trajectory $x(t)\, (t\le0),$
with continuous derivative, on which the velocities
do not exceed a certain limiting velocity $v$ smaller than the speed of light $c,$
is called an {\em initial trajectory.}
A locally Lipschitzian trajectory $y(t)$ is called {\em relativistically admissible}
if the velocities on it stay below the speed of light $c,$ though they may approach $c$
arbitrarily close.

\begin{thm}
Assume that the trajectory $y(t)\,(t\ge 0),$ extends an
initial trajectory $x(t)$ and is relativistically admissible.
Then for every such a trajectory $\ y\ $ there exist unique functions
\begin{equation*}
 t_{jk}(t)\text{ and }r_{jk}(t)\fa t\in R
\end{equation*}that appear in Feynman's formula (\ref{n-body problem})
for the intensity of the electric field $E_{jk}.$
\end{thm}


Define the  operators of a trajectory $y$ by the following formulas
\bb
    \begin{split}
        \Phi_{j}=&\ \Gamma(v_j)-[\frac{q_jq_{k}}{4\pi \epsilon_0 c^2 m_{j}}]
                        \sum_{k\neq j}H_{jk}\circ G_{jk}\qtext{ for }j=1,2,\ldots,n
    \end{split}
\ee
where $m_j$ is the relativistic mass of the charge $q_j.$
The operators $\Gamma,$ $H_{jk},$ and $G_{jk}$ have their values
in the space of linear transformations
from the Euclidean space $R^3$ into $R^3.$  They are given by the formulas
for all $h\in R^3$
\bb
    \begin{split}
    \Gamma(v_j)(h)=&\ h+ \gamma_j^2(u_j,h)u_j\\
    H_{jk}(h)=&\
    (h+\frac{1}{c}\,v_j\times\big[\,e_{jk}\times h\,\big])\\
    G_{jk}(h)=&\ |r_{jk}|^{-1}\{h+[c-(e_{jk},v_{k}(t_{jk}))]^{-1}(e_{jk},h)
    [ce_{jk}+v_k(t_{jk})]\}\\
    \end{split}
\ee
where $u_j=\frac{1}{c}v_j$ and $\gamma_j=(1-|v_j|^2c^{-2})^{-1/2}.$


A point $t=t_1$ on a trajectory $y$ is called {\em singular}
if either some of the charges collide at the time $t_1$
or the determinant is zero for at least one of the transformations
$\Phi_j(y)(t_1).$


\begin{thm}
If $x(t)\,(t\le0)$ is an initial trajectory with  nonsingular point $t=0,$
then there exists a unique relativistically admissible
trajectory $y(t),$ defined for
$t$ in an interval $I\subset \langle 0,\infty),$
extending the initial trajectory $x(t)$
and having the following properties.
\begin{enumerate}
\item No point $t$ on the trajectory $y$ is singular.
\item The trajectory $y$ represents a solution of the Newton-Einstein system of equations
with Lorentz forces induced by electromagnetic field in accord to Feynman's law for moving
point charges
(\ref{n-body problem}).
\item The trajectory $y$ cannot be extended any
further preserving the above properties that is it represents the maximal global solution
of the system
(\ref{n-body problem}).
\end{enumerate}
\end{thm}

In order to prove the theorems the author makes use of %
the theory of generalized Lebesgue-Bochner-Stieltjes
integral as developed in Bogdanowicz \cite{bogdan10} and \cite{bogdan14},
the Banach contraction
mapping theorem, and the works of Lorentz, Einstein, and Feynman.

Detailed proofs of the theorems will appear in {\em Quaestiones Mathematicae}
\cite{bogdan64}.



\end{document}